\documentclass[english]{article}
\usepackage[T1]{fontenc}
\usepackage[latin9]{inputenc}
\usepackage{textcomp}
\usepackage{amsbsy}
\usepackage{amstext}
\usepackage{stmaryrd}
\usepackage{graphicx}

\makeatletter

\providecommand{\tabularnewline}{\\}

\makeatother

\usepackage{babel}
\begin{document}

\title{Measurement Problem in Quantum Mechanics and the Surjection Hypothesis }

\author{Fritz W. Bopp\\
Universität Siegen}
\maketitle
\begin{abstract}
Starting with unitary quantum dynamics, we investigate how to add
quantum measurements. Quantum measurements have four essential components:
the furcation, the witness production, an alignment projection, and
the actual choice decision. The first two components still lie in
the domain of unitary quantum dynamics. The decoherence concept explains
the third contribution. It can be based on the requirement that witnesses
reaching the end of time on the wave function side and the conjugate
one have to be identical. In this way, it also stays within the quantum
dynamics domain. The surjection hypothesis explains the actual choice
decision. It is based on a two boundary interpretation applied to
the complete quantum universe. It offers a simple way to reduce these
seemingly random projections to purely deterministic unitary quantum
dynamics, eliminating the measurement problem. 
\end{abstract}

\section{Introduction}

One can argue that time is ripe for solving the fundamental problem
with quantum mechanics (QM) measurements. Nowadays, experimental observations
of quantum effects are available, exceeding by far what the founders
considered possible. They involve individual particles and reach in
regions that were considered a safe, purely classical refuge. QM is
a non-relativistic approximation of relativistic quantum field theory.
One managed to demystify infinities in the relevant gauge theories,
and in their domain of validity, these theories can be considered
understood. In particular, one learned that careful consideration
of what is to be taken as the final state is essential to avoid infrared
singularity.

In the last 100 years, many ideas about quantum measurements were
pursued, and there is an enormous amount of publications. They are,
unfortunately, not always of good quality and sometimes involve overstated
opinions. To proceed, we propose to divide the problem into separate
compartments. In this way, it should be clearer what addresses what
aspect. It will allow us to discuss various interpretations in a rather
neutral way and, also, to present our advocated surjection interpretation.
\pagebreak{}

Quantum mechanics (QM) contains unitary quantum dynamics and the physics
of quantum measurements. Quantum measurements can then be segmented
into four components.

\section{Quantum dynamics}

The term quantum dynamics was coined by Sakurai~\cite{sakurai2014modern}.
It means QM without measurement jumps or collapses. As Sakurai pointed
out, all the spectacular QM successes in atomic, nuclear, particle,
and solid-state physics lie in the domain of quantum dynamics. It
explains how electric fields close or open conduction zones in field-effect
transistors allowing us to compose this paper. Its field theory side
predicts, with spectacular precision, the anomalous magnetic moments
of electrons and also muons. 

With its precision and its applicability domain, it is fair to say
that it is the best-known physics field we have. It contains no fundamental
problems. The famous criticism of QM that one calculates something
that one does not fundamentally understand does not apply to quantum
dynamics. Of course, the field theory will eventually have to be adopted
at very short distances, and other changes might be attractive. However,
for the understanding of quantum measurements, this should be irrelevant.

Compelled by a quantum statistical argument~\cite{Bopp:2019opk},
we take a realist\textasciiacute s view. The wave functions or quantum
fields are the draft horses of the theory. To deny their ontological
status~\cite{oldofredi2020classification} might seem irrelevant
as long as one accepts that they can pull the plows, but this ignores
simplicity~\cite{von1955fmethod} which we consider essential. 

Quantum dynamics determines the amplitude
\begin{equation}
<i\,|\,U(t_{f}-t_{i})\,|\,f>\,.
\end{equation}
of how a given initial state evolves to a final state. Within this
expression, there are many coexisting intermediate states. They are
at least in principle calculable and knowable. Specifically, the amplitude
can be written as a sum over all contributing Feynman paths:
\begin{equation}
\sum_{j}<i\,|\,U_{j}(t_{f}-t_{i})\,|\,f>\,.
\end{equation}
The relative size of its contribution involves its absolute squares:
\begin{equation}
\begin{array}{c}
<i\,|\,U(t_{f}-t_{i})\,|\,f><i\,|\,U(t_{f}-t_{i})\,|\,f>^{*}=\\
\sum_{j,j'}<i\,|\,U_{j}(t_{f}-t_{i})\,|\,f><f\,|\,U_{j'}(t_{i}-t_{f})\,|\,i>
\end{array}
\end{equation}
where the paths on both sides are chosen independently. Except for
the initial and final states, the wave function side\textasciiacute s
choices and the complex conjugate side\textasciiacute s ones are independent.
We consider this factorization as a central property of the theory. 

If one wants to consider the contribution of an initial state $<i\,|$
to all possible final states, one has to replace the product $|\,f><f\,|$
with the unit operator $\boldsymbol{1}$ connecting the wave function
and the complex conjugate side:
\begin{equation}
\begin{array}{c}
\sum_{j,j'}<i\,|\,U_{j}(t_{f}-t_{i})\,|\,\boldsymbol{1}\,|\,U_{j'}(t_{i}-t_{f})\,|\,i>\end{array}\label{eq:unit_evolution}
\end{equation}

How does this solid quantum dynamics has to be amended to account
for measurement processes?

\section{Components of quantum measurements}

\begin{figure}[h]
\noindent \centering{}\includegraphics[scale=0.4]{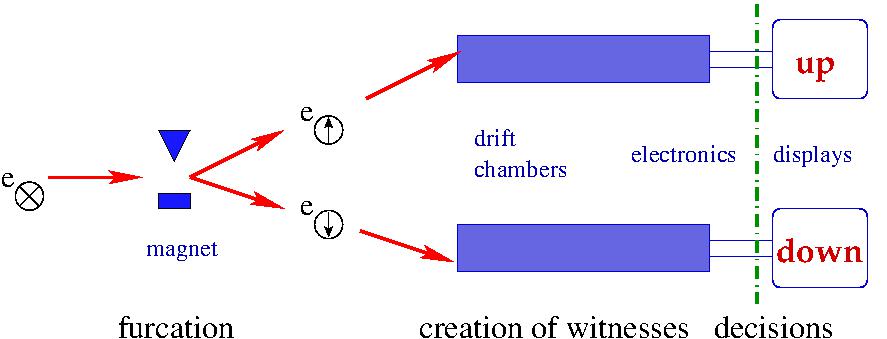}\caption{Stern-Gerlach measurement}
\end{figure}
We turn to an idealized Stern-Gerlach arrangement shown in figure~1.
The wave function of an electron with a spin showing into the screen-plan
gets separated in an inhomogeneous magnetic field into its up and
down components both entering distinct drift chambers in which some
electrons are knocked of their atoms and collected by a charged coupled
electronics flushing ``up'' or ``down'' on display. 

The process involves four essential parts: 

\noindent \begin{center}
\begin{tabular}{|c|}
\hline 
furcation\tabularnewline
\hline 
witness production\tabularnewline
\hline 
alignment projection\tabularnewline
\hline 
choice decision\tabularnewline
\hline 
\end{tabular}
\par\end{center}

\noindent which will be explained below. They appear in all quantum
measurements. \pagebreak{}

\section{Furcation and witness production}

At the furcation, the wave functions are split. The separation need
not be geometrical; it must just enable the distinction in witness
production in the next step. The furcation time can be much earlier
than this next step. Furcated states (involving entangled partners)
were shown to travel more than a thousand kilometers~\cite{yin2020entanglement}.

A vast number of low-energy photons typically dominate the witness
production step. Most devices emit measurable Radio-frequency waves.
Such photons with a wavelength between $0.03$ and $20$ meters carry
unmeasurable energy of $10^{-23}$and $10^{-26}$ Joule, which means
there are a lot of them. They are penetrating and have an excellent
chance to escape the apparatus, the walls, and the ionosphere escaping
into the open space. 

For an expanding universe, it means they live practically forever.
For simplicity, we will take the life-time of the universe, $t_{f}$,
as finite. Presumably, in an eventually no longer interacting thin
universe, the limit, $t_{f}\to\text{\ensuremath{\infty}}$, would
exist. 

In the setup of figure~1, the choice of the emitting drift chamber
is in this way encoded in the hugely extended final state. It remains
traceable. One can evolve the final state backward. As the photon
number is large, one can disregard all components which interact before
the drift chambers are reached. The procedure is not disturbed by
the massive number of other witnesses. 

In the quantum-dynamical evolution, many splittings and mergings contribute.
We assume that this encoding of a splitting choice in the final state
is a defining property of what has to be considered a quantum measurement. 

To conclude, both processes, i.e., the furcation and the witness production,
are still in the domain of quantum dynamics. However, they are essential
ingredients of the measurement process. The measurement\textasciiacute s
actual decisive part then has two parts: the alignment projection
and the seemingly-random-choice decision.

\section{Alignment decision }

\noindent In QM, all combinations of components of the wave function
and their conjugate coexist. The first reduction in the measurement
decision is the alignment projection which selects matching components
and eliminates mixed terms. Consider the up and down ones of the above\textquoteright s
setup. The alignment decision here introduces the following projection
operator:
\begin{equation}
\;\begin{array}{ccc}
P_{\mathrm{no\,interference}}(t) & = & |\uparrow>^{\mathrm{wave}}<\uparrow|\mathrm{^{wave}}\hspace{0.3cm}\otimes\hspace{0.3cm}|\uparrow>^{\mathrm{conjugate}}<\uparrow|\mathrm{^{conjugate}}\\
 & + & |\downarrow>^{\mathrm{wave}}<\downarrow|\mathrm{^{wave}}\hspace{0.3cm}\otimes\hspace{0.3cm}|\downarrow>^{\mathrm{conjugate}}<\downarrow|\mathrm{^{conjugate}}
\end{array}
\end{equation}

This projection operator connects the wave function world and its
conjugate. To illustrate how this happens, one can write equation
4 so that the part before the unit operator goes upward and the part
after it goes downward, as done in equation 6. In this way, the vertically
upward direction corresponds to the physical time direction denoted
by $t$. The horizontal direction corresponds to the quantum operations
order denoted as ``quantum time,'' by $\tau$. 

\noindent \begin{center}
\includegraphics[bb=4bp 14bp 423bp 165bp,clip,scale=0.5]{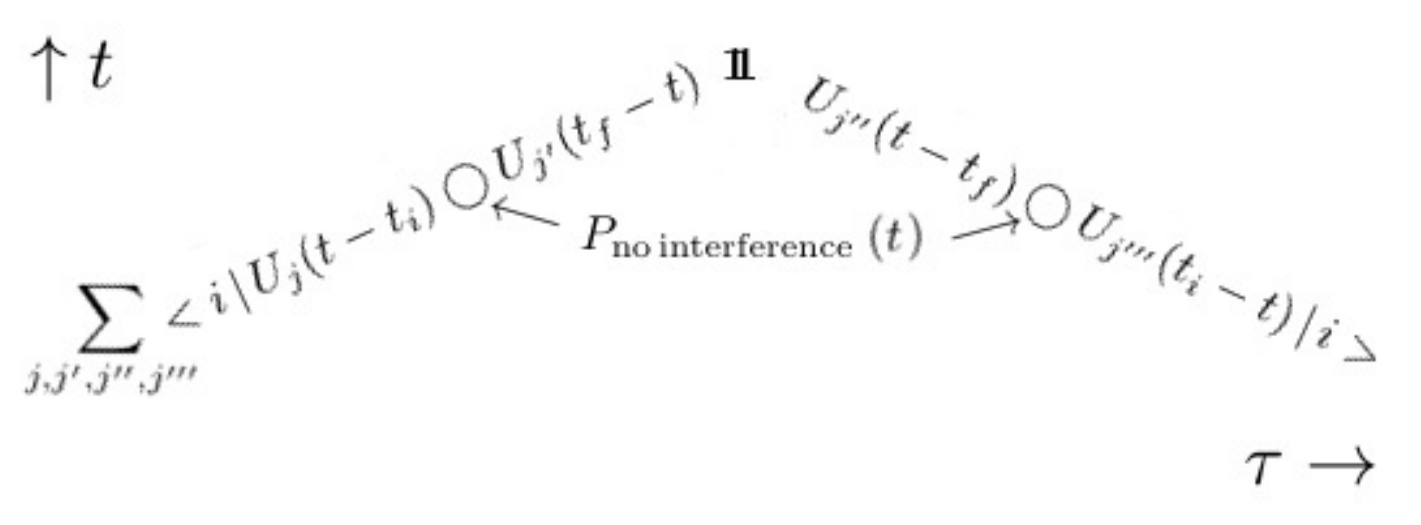}
\begin{equation}
\,
\end{equation}

\par\end{center}

\noindent The projection operator which eliminates interference terms
connects both sides. Both sides are otherwise not linked, and this
simple addition seems to violate the spirit of a quantum dynamic-based
theory. 

In Bohmian mechanics~\cite{bohm1952suggested,durr2009bohmian}, the
alignment follows from the assumption that the guided point particle
does not live in wave function/conjugate quantum world and has to
have one of the two values. Such an intrinsic automatic alignment
also seems to apply to hidden variable theories~\cite{bell1964einstein,Hooft:2016Cellular}. 

In objective-collapse theories~\cite{ghirardi1986unified,Davidson:2020fpx}
the situation is more complicated. A new physical jump process introduces
a projection operator:
\begin{equation}
P_{\mathrm{jump}}(t)=|\uparrow><\uparrow|\,\,\,\,\,+\,\,\,\,\,|\downarrow><\uparrow'|
\end{equation}
in the world or its complex conjugate. In equation 7, the meaning
of $<\uparrow'|$ is not clear. If one would assume $<\uparrow'|\uparrow>\ne0$,
a first-order process in jump dynamic would exist, but $<i\,|\downarrow>^{\mathrm{world}}P_{\mathrm{jump}}<\uparrow|\,i>^{\mathrm{conjugate}}$
would introduce an angular dependence not seen in experiments. A second-order
interaction is required acting on the wave function and its conjugate
at the same time. The violation of the above separation postulate
is therefore not avoided. Also, such new processes do require a relevant
scale. As interference effects are observed at an astronomical scale~\cite{brown1957interferometry},
it is hard to explain how an objective-collapse theory could be used
to understand table-top experiments. 

Most seriously, in these theories or in a theory that adds the operator
given in equation 5, it is not understood why witnesses\textasciiacute{}
production part is an essential ingredient of the measurement process.
Therefore, we feel compelled to reject them. 

It is widely agreed that decoherence~\cite{joos2013decoherence}
offers a simple mechanism to achieve the needed projection by eliminating
interference contributions. The idea relies on that everything including
witnesses from both sides have to agree. Traditionally witnesses are
considered to reach an outside ``macroscopic'' domain where both
components can no longer coexist separately and therefore have to
coincide. However, there is not enough justification for adding such
an extra outside domain as it can easily be avoided. 

Let us look at the many world interpretation (MWI)~\cite{vaidman2014many}
in a realist\textasciiacute s way. Each measurement decision results
in a choice of separate worlds. It is not the furcation part but the
witness production process which starts ``new worlds.'' A defining
property of a separate world is that it can no longer merge with other
worlds as different witnesses prevent such interference from different
worlds for all future times, at least ``for all practical purposes.''
The outside macroscopic domain is not needed. It is replaced by some
kind of boundary condition involving the witnesses.

Let us consider the way this alignment mechanism arises in a finite
universe. Not requiring specific final properties replaces the final
density matrix by the identity operator as it was done in equation
4. The identity operator then also involves the witnesses and requires
them to match as pictured in equation~8:

\noindent \begin{center}
\includegraphics[bb=14bp 20bp 423bp 310bp,clip,scale=0.5]{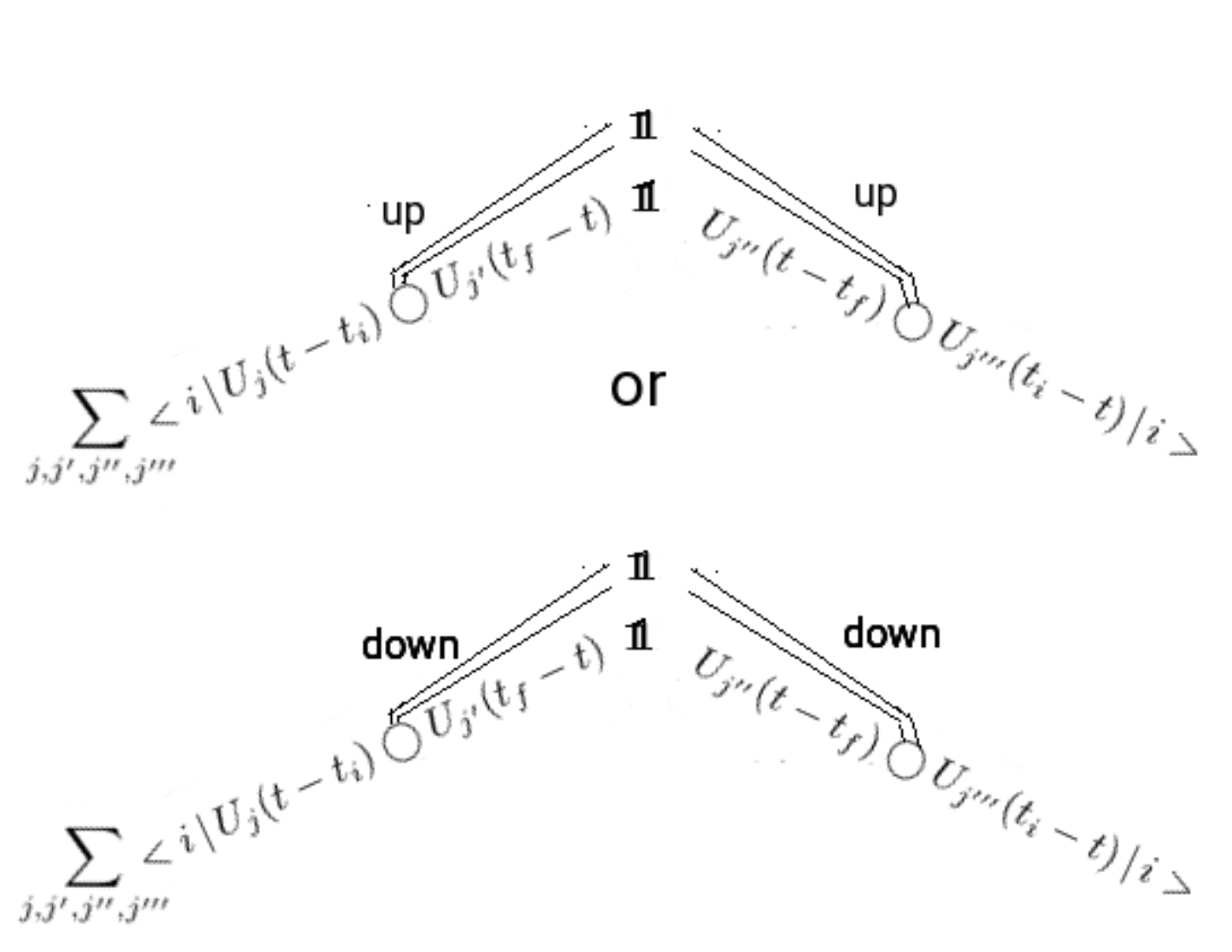}
\begin{equation}
\,\,\,\,\,\,\,\,\,\,\,\,\,\,\,\,\,\,\,\,\,\,\,\,\,\,\,\,\,\,\,\,\,\,\,\,\,\,\,\,\,\,\,\,\,\,\,\,\,\,\,\,\,\,\,\,\,\,\,\,\,\,\,\,\,\,\,\,\,\,\,\,\,\,\,
\end{equation}

\par\end{center}

\noindent which effectively creates the required projection. The concept
is attractive as no departure of quantum dynamics was required. 

As a side remark, the concept contains some indirect backward causation
on a wave function level. Decisions on the wave-function side determine
the final state at $t_{f}$. The matching at $t_{f}$ introduced a
projection affecting the complex conjugate wave function at much earlier
times. However, backward causation on a quantum level does not exclude
effective causality on a macroscopic level~\cite{Bopp:2016nxn}.

\section{Seemingly-random-choice decision}

The remaining projection implements the seemingly random choice between
the different possible states. It is a central problem of quantum
measurements, and many concepts were proposed. The most common implementation
is just a randomly chosen projection operator in the wave function
evolution or its conjugate 
\begin{equation}
P_{\mathrm{jump}}(t)=|\uparrow><\uparrow|\,.
\end{equation}
with a probability reflecting the size of the amplitude-squares. The
introduction of the projection operator in the evolution

\begin{equation}
\begin{array}{c}
<i\,|\,U(t-t_{i})\,P_{\mathrm{jump}}(t)\,U(t_{f}-t)\,|\,\boldsymbol{1}\,|\,U(t_{i}-t_{f})\,|\,i>\end{array}
\end{equation}
can be written in a symmetric way relying on the alignment mechanism
of the witnesses 

\begin{equation}
\begin{array}{c}
<i\,|\,U(t-t_{i})\,P_{\mathrm{jump}}(t)\,U(t_{f}-t)\,|\,\boldsymbol{1}\,|\,U(t-t_{f})\,P_{\mathrm{jump}}(t)\,U(t_{i}-t)\,|\,i>\,.\end{array}
\end{equation}
In this expression, the projections mainly affect the future $t'>t$~.
However, the projection can also eliminate paths in the past. It is
experimentally known that for entangled states, such backward correlations
appear on a wave function level.

The Copenhagen interpretation tries to avoid backward causation by
denying wave functions and fields ontological reality, which opens
the door for intricate philosophical complications~\cite{friebe2015messproblem}.
It seems largely successful except for quantum statistical effects,
which introduce backward causation on a particle and not just on a
wave-function level~\cite{bopp2001ulm,Bopp:2016nxn,bopp2019bi,Bopp:2019opk}.
The denial might have too much respect for classical physics and its
concepts. Quantum mechanics with fields and wave functions is a beautiful
well-tested theory with contains backward causation. The only requirement
is that classical physics, with its seemingly causal behavior, emerges
approximately to the degree and where we know it~\cite{Bopp:2016nxn}.

An intricate point concerns the probability with which the chosen
contribution appears. The statistical relative weight of the decision
at the time $t$ depends on on the wave function and its conjugate:
\begin{equation}
\frac{\begin{array}{c}
<i\,|\,U(t-t_{i})\,P_{\mathrm{jump}}^{\uparrow}(t)\,U(t_{f}-t)\,|\,\boldsymbol{1}\,|\,U(t-t_{f})\,P_{\mathrm{jump}}^{\uparrow*}(t)\,U(t_{i}-t)\,|\,i>\end{array}}{\begin{array}{c}
<i\,|\,U(t-t_{i})\,P_{\mathrm{jump}}^{\downarrow}(t)\,U(t_{f}-t)\,|\,\boldsymbol{1}\,|\,U(t-t_{f})\,P_{\mathrm{jump}}^{\downarrow*}(t)\,U(t_{i}-t)\,|\,i>\end{array}}\,.
\end{equation}
It means that both sides are needed. Our quantum-dynamics-based postulate
does not allow such a dependence except for the endpoints. 

As said, in our epoch in the universe, typical witnesses live practically
forever. Hence, the time - when the projection operator acts - is
irrelevant. An obvious way to adhere to our postulate is to choose
$t=t_{f}$ and we write for the probability of a particular jump choice:

\begin{equation}
\mathrm{prob(jump)}=\frac{<i\,|\,U(t_{f}-t_{i})\,P_{\mathrm{jump}}(t_{f})\,|\,\boldsymbol{1}\,|\,P_{\mathrm{jump}}^{*}(t_{f})\,U(t_{i}-t_{f})\,|\,i>}{<i\,|\,U(t_{f}-t_{i})\,|\,\boldsymbol{1}\,|\,U(t_{i}-t_{f})\,|\,i>}
\end{equation}

The availability of surviving witnesses at $t=t_{f}$ determining
the choice $P_{\mathrm{jump}}(t_{f})$ is, as said, taken as the definition
of a true measurement process. There are many such branching processes
in the evolution of the universe. They are defining a path, $Path_{k}$,
in a huge tree in a many-worlds like structure. As there is no overlap
between witness configurations, the path has to be equal on both sides,
and one can write:
\begin{equation}
\sum_{\mathrm{all}\,k'}<i\,|\,U(t_{f}-t_{i})\,Path_{\mathrm{k'}}(t_{f})\,|\,\boldsymbol{1}\,|\,Path_{\mathrm{k'}}^{*}(t_{f})\,U(t_{i}-t_{f})\,|\,i>=\boldsymbol{1}
\end{equation}

\noindent where $Path_{\mathrm{k'}}(t_{f})=P_{\mathrm{jump(1)}}^{k'}(t_{f})\,\,\cdots\,\,\cdot P_{\mathrm{jump(n_{i})}}^{k'}(t_{f})\,$
and the conjugation just changes the ordering. 

The random choice decision has to select a given $path_{k}$ with
a probability of the weight of the path
\begin{equation}
\mathrm{prob}(k)=<i\,|\,U(t_{f}-t_{i})\,Path_{\mathrm{k}}(t_{f})\,|\,\boldsymbol{1}\,|\,Path_{\mathrm{k}}^{*}(t_{f})\,U(t_{i}-t_{f})\,|\,i>\label{eq:prob(k)}
\end{equation}
A computer program would first divide the state on both sides to mutually
orthogonal path contribution, calculate their weight $\mathrm{prob}(k)$,
and then choose a random number $RAND\in[0.1]$ and take the largest
$k$ with $\sum_{k'=1}^{k}\mathrm{prob}(k')<RAND$ to select the chosen
$path_{k}$. It is not a trivial task. The process should not affect
self-organizing processes, which play an essential role in the evolution
of the universe.

The obtained description is usually called two boundary
quantum mechanics~\cite{hartle2020arrows,wharton2010time,Wharton:2011kw}.
It is close to a multi-world interpretation in which the path is determined
by a community of observers seeing identical measurement results.
The probability of such a community witnessing the $path_{k}$ is
determined analogously using equation~\ref{eq:prob(k)}. The matrix:
\begin{equation}
M_{\mathrm{all\,jumps}}(t_{f})=Path_{\mathrm{k}}(t_{f})\,|\,\boldsymbol{1}\,|\,Path_{\mathrm{k}}(t_{f})
\end{equation}
is a hermitian with extremely tiny uncorrelated eigenvalues. This
situation might allow approximating the matrix by the dominant vector
product:
\begin{equation}
M_{\mathrm{all\,jumps}}(t_{f})=|t_{f}^{\mathrm{dominant}}><t_{f}^{\mathrm{dominant}}|\,.
\end{equation}
In this way, one would obtain the two-state-vector interpretation
of Aharonov and collaborators~\cite{aharonov1964time,aharonov2017twotime}. 

The two-boundary and the two-state-vector descriptions are valid super-deterministic
interpretations of QM~\cite{hossenfelder2020rethinking}. However,
the needed procedure to obtain equation~\ref{eq:prob(k)} without
affecting self-organization processes might be too complicated to
accept. Also, such interpretations are incompatible with free will
concepts.

There is a simple, natural way to select a path according to equation~\ref{eq:prob(k)}~\cite{bopp2019bi},
avoiding random selection. If desired, it can be easily amended to
include free will decisions~\cite{bopp2020avoid}.

Consider the evolution of a state with a particular spin choice, say
spin up, i.e., $<\uparrow(t)_{i}\,|$, in a universe originating in
the initial state $<i\,|$~. If all final state $|f_{k}>$ are accepted
unitarity leads to:
\begin{equation}
\sum_{\mathrm{all\,}k}<\uparrow(t)_{i}\,|\,U(t_{f}-t)\,|\,f_{k}>=1\,.\label{eq:evolution_tro_f}
\end{equation}
The probability of the up spin state choice reflects the production
not considered in equation \ref{eq:evolution_tro_f}.

If the initial state on the wave function side $|\,i>$ differs from
that on the conjugate side $|\,i'>$ equation \ref{eq:unit_evolution}
changes to: 
\begin{equation}
<i\,|\,U(t_{f}-t_{i})\,\boldsymbol{\boldsymbol{1}}\,U(t_{i}-t_{f})\,|\,i'>\,\,\,\,\,\,\,\ll\mathrm{1}\,\,\,.\label{eq:overlapp}
\end{equation}
Considering the final universe\textasciiacute s rich structure, only a small
region in the vast total Hilbert space involves each side\textasciiacute s
final state. If the state $|\,i'>$ is unrelated to $|\,i>$, the
overlap (equation \ref{eq:overlapp}) can be expected to be tiny by
many orders of magnitudes. For equation \ref{eq:evolution_tro_f},
it means:
\begin{equation}
\sum_{k\in\mathrm{overlapp}}<\uparrow(t)_{i}\,|\,U(t_{f}-t)\,|\,f_{k}>\,\,\,\,\,\ll\mathrm{tiny}\,\,\,.
\end{equation}
The same holds for the down state.

In the considered experiment, we started at the time $t$ with a prepared
in-the-screen-spin state $<\varotimes{}_{i}\,|\,$. The probability
of measuring an up-spin state involves the absolute square of the
amplitude:
\begin{equation}
\left|<\varotimes{}_{i}\,|\uparrow(t)_{i}>\sum_{k\in\mathrm{overlapp}}<\uparrow(t)_{i}|\,U(t_{f}-t)\,|\,f_{k}>\right|^{2}.
\end{equation}
 Statistically, two unrelated, extremely tiny quantities can not have
a comparable magnitude, and one choice will dominate:

\begin{equation}
\begin{array}{ccc}
\left|<\varotimes{}_{i}\,|\uparrow_{i}>\begin{array}{c}
\sum\\
k\in\\
\mathrm{overlapp}
\end{array}<\uparrow_{i}|U|f_{k}>\right|^{2} & \begin{array}{c}
\gg\\
\mathrm{or}\\
\ll
\end{array} & \left|<\varotimes{}_{i}\,|\downarrow{}_{i}>\begin{array}{c}
\sum\\
k\in\\
\mathrm{overlapp}
\end{array}<\downarrow_{i}|U|f_{k}>\right|^{2}.\end{array}
\end{equation}
 It will look like a random decision. 

Averaging over many observations  in different situations, one obtains: 
\begin{equation}
\left[\left|\sum_{k\in\mathrm{overlapp}}<\uparrow_{i}|U|f_{k}>\right|^{2}\right]=\left[\left|\sum_{k\in\mathrm{overlapp}}<\downarrow_{i}|U|f_{k}>\right|^{2}\right]\,.
\end{equation}
Moreover, the transition at a time t is independent of the environment.
Hence, the Born term: 
\begin{equation}
\left|<\mathrm{\varotimes}|\uparrow>/<\mathrm{\varotimes}|\downarrow>\right|^{2}
\end{equation}
determines the relative contribution.

The obtained quantum decision is in some way like in a hidden variable
theory. The ``hidden'' variable $\left(|\,i>-|\,i'>\right)$ is,
however, not affixed to individual observed particles but involves
the whole universe. It determines the overlapping final density matrix,
which then fixes the path of the measurement choices. 

The presented concept also allows for this last step to stay in quantum
dynamics. It is, therefore, without internal inconsistencies usually
associated with measurements. Random decisions are gone, which is
of fundamental importance. They were the reason why Einstein could
not accept QM~\cite{einstein1935can} as complete.

\section{The surjection view of the evolution}

\begin{figure}[h]
\noindent \centering{}\includegraphics[scale=0.3]{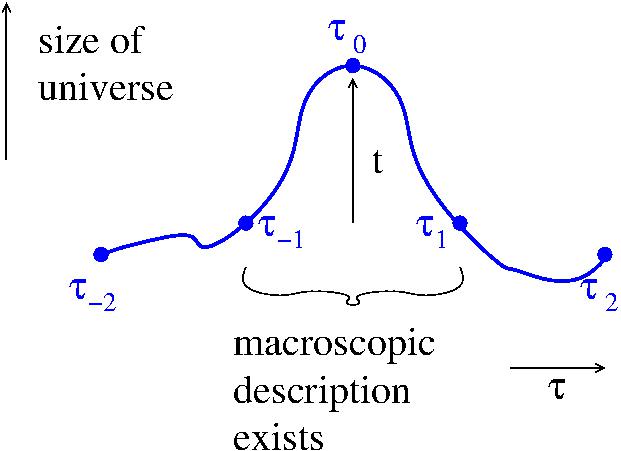}\caption{The surjection concept}
\end{figure}

Our interpretation is based on a surjection concept which changes
our view of cosmological evolution. It considers a quantum universe
starting, say, at a quantum time $\tau_{\ensuremath{-2}}$ reaching
a huge maximal extension at a quantum time $\tau_{\ensuremath{0}}$
and ending at a quantum time $\tau_{\ensuremath{2}}$, as illustrated
in figure~2. The quantum times $\tau_{\ensuremath{-1}}$ and $\tau_{\ensuremath{1}}$
lie in the expanding and contracting phase with an equal distance
to the zenith, i.e. $(\tau_{0}-\tau_{-1})=(\tau_{1}-\tau_{0})$. Usually,
no macroscopic, classical description is possible. An exception is
a region between the quantum-times $\tau_{\ensuremath{-1}}$ and $\tau_{\ensuremath{1}}$.
It involves a surjective mapping from a wave function $\psi(\tau)$
with $\tau\in[\tau_{-1},\tau_{0}]$ and a wave function $\psi(2\tau_{0}-\tau)$
with $(2\tau_{0}-\tau)\in[\tau_{0},\tau_{1}]$ to macroscopic object
$\psi(2\tau_{0}-\tau)\cdot\psi(\tau)\approx\psi^{*}(t)\cdot\psi(t)$
with the real-time $t$ a monotonic function of $\tau-\tau_{-1}$.
In simple words, we live with our wave function in the expanding quantum
world and with our conjugate one in the contracting one. 

Close to the zenith, one has $\psi(\tau_{0}+\epsilon)=\psi^{*}(\tau_{0}-\epsilon)$
. The quantum-times $\tau_{\ensuremath{-1}}$ and $\tau_{\ensuremath{1}}$
are chosen in a way that in the in-between region, the extensive system
of witnesses matching in the extremely extended region at $\tau_{0}$
keep $\psi(2\tau_{0}-\tau)$ and $\psi^{*}(\tau)$ sufficiently close
to maintain the usual prediction. Eventually, the different ``initial''
states
\[
<i\,|=<\psi(\tau_{-2})\,|\mathrm{\,\,and\,\,}|\,i'>=|\psi(\tau_{2})>
\]
will yield $\psi(2\tau_{0}-\tau)\ne\psi^{*}(\tau)$ leading to visible
CPT violations~\cite{bopp2020avoid}. 

The required overlap at $\tau_{0}$ biases the contributing states
at $\tau_{1}$ and $\tau_{-1}$. A more homogenous state at $\tau_{1}$
will at $\tau_{-1}$ have a larger domain close by with sufficiently
matching partners than one with unregular structures. The - in this
way - favored homogeneity in the early universe is observed. It is
usually attributed to inflation~\cite{guth1981inflationary}.

\section{Conclusion}

Most of the spectacular successes of QM lie in the domain of quantum
dynamics. We investigate how to add measurements and define four components.
We observe that both setup components actually lie in the domain of
quantum dynamics. The third component is the elimination of interference
terms of components corresponding to different measurement results.
The decoherence argument explains this contribution. If the argument
is based on witnesses reaching the final state, it also manages to
stay within the quantum dynamics domain. 

We argue for a simple way to also reduce the seemingly random jump
projections to the pure deterministic quantum dynamics. The universe\textasciiacute s
lifetime is assumed to be finite with an open matching state at the
final real-time. Two boundary states remain, the initial state on
the wave function side and that on the usually just conjugate side.
They are here taken as unrelated. Their difference acts as a hidden
variable fixing a narrow matching state at the final time which then
determines the seemingly random quantum choices. In this way, quantum
measurements are embedded in entirely consistent quantum dynamics. 

\bibliographystyle{plain}
\bibliography{literatur}

\end{document}